\documentclass[preprint,amsmath,amssymb,11pt,english,aps,showpacs,showkeys,pra]{revtex4-2}
\usepackage{graphicx}
\usepackage{graphics}
\usepackage{amsmath}
\usepackage{dcolumn}
\usepackage{amssymb}
\usepackage{bm}
\usepackage[utf8]{inputenc}

\begin{document}
\title{Quantum Holonomies in Graphene Wormholes}
\author{Everton Cavalcante}
\email{Electronic address: everton@servidor.uepb.edu.br}
\affiliation{Departamento de Física, Centro de Ci\^encias e Tecnologia, Universidade Estadual da Para\'{\i}ba, Campina Grande, PB, Brazil}


\begin{abstract}

In this paper, a discussion about quantum holonomies around a possible bridge between two graphene sheets has been made. That bridge is widely known as a graphene wormhole, and some of its characteristics are also showed up here. As well as their build as a zigzag junction between a baggy nanotube and the graphene lattices. And how the localized electronic states could be mimicked by gauge fields in a low-energy regime. Further, the possibility to build holonomies handle by an effective flux from topological defects in junctions of that bridge has been discussed.

\end{abstract}



\maketitle

\section{Introduction}\label{secI}
In the '80s some works about low-dimensional physical systems were on rising \cite{Jackiw, Deser Jackiw}. In addition to having an excellent pedagogical role, the gravitational ones had been important as a precursor to models in condensed matter physics. The most revolutionary one is graphene. Many of the outstanding features of graphene arise from the offbeat band structure of the carbon sheet. Instead of the gaped demeanor of conduction (and valence) band from undoped materials, this one has a conical meet at the corners of the hexagonal Brillouin zone. Thus mimicking the dispersion relationship from fermions in the Dirac (or Weyl) equation at low energies. Furthermore, a quasiparticle on graphene has a relativistic behavior, although it is not really a relativistic particle \cite{Katsnelson}. 
The fact that, at low energies, fermions in graphene still preserve their relativistic invariance is the origin of many remarkable properties as, for example, their insensitivity of the electronic transport to scatterers with a size large than the lattice spacing \cite{Suzuura}.

Another interesting aspect of graphene concerns its curvature. It is widely known that samples of them always presented some corrugation \cite{Fasolino}. Nonetheless, most of these corrugations cannot be associated with substrate imperfections. It's deeper than that! Suspended samples also have corrugation. In the reference \cite{Katsnelson Geim Philos}, Katsnelson discusses the relationship between these ripples and how a fictitious gauge field could be simulated for electrons hooping across atoms. 

A further interesting effect is the presence of topological defects in graphene. Taking into account is at least very difficult to build a real sample without them. A famous approach to topological defects in materials has been made for Volterra. Widely called of "cut and glue" process \cite{Volterra}. Moreover, a geometrical description for disclinations and dislocations in solids was presented by Katanaev and Volovich \cite{Katanaev and Volovich}. For that, a Riemann-Cartan geometry was claimed, and these topological defects could be mimicked by a non-Abelian gauge field source. Thus we can confer to graphene the status of a possible laboratory for field theory in (2+1)-dimensional systems in the topological defect's background \cite{Fialkovsky Vassilevich}.

In fact, the presence of curvature on graphene could be strong enough to change its own topology. And, one of the most remarkable possibilities is the graphene wormholes. Over a large scale is expected some non-zero curvature on the graphene sample. It does mean the appearance of non-usual carbon rings. In this way, we can imagine two possibilities right away: pentagon rings and heptagon rings. The first of them is related to the build of fullerenes \cite{Kroto, GonzalezGuineaVozmediano1, GonzalezGuineaVozmediano2, Everton Claudio1, Everton Claudio2}. The second one is related to graphene wormholes \cite{Gonzalez Herrero, Garcia Furtado}. Both of them need 12 unusual rings to be created. That is a striking resemblance!

In the general relativity (GR) scenario, the first wormhole solution was put forward by Einstein \cite{Einstein Rosen} as a possible bridge that was made possible by a coordinate change in Schwarzschild geometry. After that, many works were developed on the subject. A precious review was made in \cite{Morris Thorne}. Nonetheless, when we have been talking about wormholes, it is not necessary to worry about energy conditions in GR. In fact, graphene wormholes must not satisfy such conditions. These kinds of wormholes were built for two graphene-nanotube junctions which play the hole of asymptotical flat spaces connected through a carbon nanotube with a zig-zag boundary mimicking a bridge \cite{Gonzalez Herrero, Gonzalez Guinea Herrero}. 
To avoid the effects of soft variations in the junctions of the bridge, it is been assumed a hole with a large radius. Much bigger than the throat length. Actually, recently some papers have shown the radius of the throat must be closer than $70 \AA$ \cite{Elllis Bronikov Ramos Furtado, catenoid bridge}. Which represents around 50 times the usual interatomic distance in the lattice ($d \approx 1,42 \AA$) \cite{Katsnelson}.

González and Herrero have still shown \cite{Gonzalez Herrero} if graphene wormholes have maximal gauge flux, the localized states would be arranged into two triplets that are the counterpart of the zero-energy modes of the fullerene lattices in the continuum limit. It is stunning that zero-energy modes of Weyl equation could be arranged in this way in the continuum model. 
Furthermore, the Dirac quantum field approach for electrons in graphene samples in the presence of topological defects is described by local frames, and the spinors transform under infinitesimal Lorentz transformations \cite{Kleinert Gauge Fields in Condensed Matter}. 
The local Lorentz group will be unchanged if a covariant derivative $\nabla_{\mu}=\partial_{\mu}+\Gamma_{\mu}(x)$ has been introduced. This allows describing a geometric phase on spinor as a function of $\Gamma_{\mu}(x)$ \cite{Bohm Mostafazadeh}.

One of the most remarkable applications for geometric phases is the possibility to build a holonomic quantum computation (HQC) \cite{Zanardi Rasetti} starting from these non-Abelian ones. In nowadays called Quantum Holonomies. They use to be stable \cite{Kuvshinov Kuzmin}. Which makes it easier the application in a wide kind of systems, such as ion traps \cite{Monroe Meekhof King Itano Wineland, Pachos Topological features in ion-trap holonomic computation}, 
Josephson-junction devices \cite{Cholascinski Quantum holonomies with Josephson-junction devices}, neutral atoms \cite{Recati Holonomic quantum computation with neutral atoms}, optical computer \cite{Pachos Optical holonomic quantum computer}, conical defects in graphene layers \cite{Bakke Holonomic quantum computation associated with a defect structure of conical graphene}, nuclear magnetic resonance \cite{Jones Geometric quantum computation using nuclear magnetic resonance} and nuclear Zeeman-perturbed systems \cite{Teles Experimental implementation of quantum information processing by Zeeman-perturbed nuclear quadrupole resonance}. Recently quantum holonomies have been studied around topological defects in graphene mimicked by an extra-dimensional approach \cite{Everton KK}.    
My aim in this work is to discuss a way to get a holonomy matrix in graphene wormholes as proposed through a Dirac phase factor method. As well as discussing the behavior of the holonomy in the neighborhood of wormhole throat. In addition to a possible practical end of the model in Josephson-junctions devices.

This paper has been organized as follows. In section \ref{secII}, a brief review of the general aspects of the problem has been presented. The most suggestive issues about graphene wormholes geometry have been brought out. As topological defects acting as gauge fields in the continuum limit, and the relation between the throat and radius of them. In section \ref{secIII}, I study the geometrical phases of the system and I share a way to obtain the quantum holonomies for massless Dirac fermions. Finally, in section \ref{secIV}, the conclusions, and possible applications, have been presented.

\section{Graphene Wormhole Geometry}\label{secII}

It's widely known graphene is a carbon honeycomb framework, made as a superposition of two triangular sublattices (indexed by $A$ and $B$). 
To the four valence electrons in each carbon vertice, three of them imply an elasticity of the structure ($\sigma$-bonds), as long the last one ($\pi$-bonds) is responsible for electronic features. 
Moreover, the spectrum of the structure is brought to arise considering the electron hopping probability between neighbors, as a known tight-binding model. And the energy band could be found using the unit cell representation in momentum space, called the first Brillouin zone. When the electrons are close to momentum Fermi points ($K$ and $K'$), we have a low excitation regime, and the relativistic behavior of the system arises as a Weyl equation: $H=-\hbar \nu_{f}\vec{\sigma}.\vec{k}$. Written by the Fermi velocity ($\nu_{f}$) and the Pauli matrices ($\vec{\sigma}$) \cite{Cristina Bena}.

Concerning the graphene wormhole description, it is a must highlight how the presence of topological defects disturbs the graphene behavior. First of all, a graphene wormhole is not a usual gravitational wormhole. They consist of a short nanotube bridging two graphene sheets. To be precise, they're two zig-zag junctions connected in the nanotube extremities \cite{Gonzalez Herrero}.
One important point to consider might be these kinds of arrangements imply the existence of six heptagon rings at each junction \cite{Gonzalez Guinea Herrero}.  And a further issue is: we must consider the radius of the role much larger than the nanotube length, so we throw away soft curvature effects. Actually, as far as possible I'll handle the null curvature out of the junction.  
From a practical point of view, we just must consider the ratio as a few dozen more than the interatomic CC-bond. As is shown in \cite{Elllis Bronikov Ramos Furtado, catenoid bridge}.

For graphene wormholes geometry, it's useful to point out two polar coordinates charts covering the lower ($r_{-}, \theta_{-}$) and upper ($r_{+}, \theta_{+}$) sheets. Both of them the respective metrics: $g^{\pm}_{ab}=diag(1,r_{\pm}^{2})$. Assumed $R$ represents the throat of the wormhole, the coordinates are related as:

\begin{equation}
r_{-}=\frac{R^2}{r_{+}}, \quad \text{with} \quad r_{+} \ge R.
\end{equation}
However, one easier way might be adopted to cover the whole wormhole using just one chart \cite{Matt Visser}: 

\begin{equation}
ds^2=dr^2+r^2d \theta^2; \qquad \textrm{for} \quad r \ge R,
\label{outside}
\end{equation}

\begin{equation}
ds^2=\bigg ( \frac{R}{r} \bigg )^4  ( dr^2+r^2d \theta^2 ); \qquad \textrm{for} \quad r \le R.
\label{inside}
\end{equation}

At this point, it seems appropriate to discuss a short overall concerning the geometrical properties of a  general wormhole. 
One of the first problems wormholes bring out is their instability in an electromagnetic context \cite{Misler Wheeler}. Such instability can be avoided if the metric has some preconditions. Such as throat and flare-out requirements \cite{Elllis Bronikov Ramos Furtado, Ellis, Bronnikov, Morris Thorne 2}.

For a wormhole to remain stable and allow fermions through the throat, it is necessary that both: the flare-out condition $ \bigg (\frac{r}{b(r)} -1> 0 \bigg )$, and throat condition $\bigg ( \frac{b(r)-\dot{b}(r)r}{2b(r)^2} >0 \bigg )$  be satisfied for a  metric of a general wormhole \cite{Non violation conditions Godani, Stability thin-shell Godani}:

\begin{equation}
ds^2=e^{2\phi(r)}dt^{2} - \frac{dr^2}{1-\frac{b(r)}{r}} - r^{2}d\theta^{2} - r^{2}\sin^{2}\theta d\phi^{2}.
\end{equation}

Where $\phi(r)$ means the redshift term. That is not useful in our static approach ($\phi(r), dt \to 0$). And $b(r)$ means the shape function. Furthermore, that model is backgrounded in a $d\phi \to 0$ symmetry.

It's easy to check both conditions are satisfied for both metrics: (\ref{outside}) and (\ref{inside}). 
As far as such metrics are regular everywhere, it's possible to introduce the Heaviside step function ($\Theta(r)$) by a conformal $\Omega(r)$ factor as:

\begin{equation}
ds^2=\Omega^2 (r) ( dr^2+r^2d \theta^2 ), \qquad \Omega (r)= \bigg ( \frac{R}{r} \bigg )^2 \Theta (R-r)+\Theta (r-R),
\label{geometric description}
\end{equation}
where the metric tensor can be rewritten by
\begin{equation}
g_{\mu \nu}=\Omega^{2}(r)\left(\begin{array}{cccc}1 & 0 \\ 0 & r^2 \end{array}\right)
\end{equation}

At this time, a formalism should be chosen to go ahead. According to the problem shown, the Cartan one seems more reasonable. Both because it has already been chosen by most part of works with massless fermions, so because their co-frame change it's usually easier to work. Here, the co-frame association ($\theta^{A}={e^{A}}_{\mu}(x)dx^{\mu}$) is made as:
\begin{equation}
\theta^r =\Omega (r)dr, \qquad \theta^\theta =d\theta.
\end{equation}
Where the diagonal tensor ($\delta_{AB}$) rewrites the metric by: $ds^{2}=\delta_{AB}\theta^{A}\theta^{B}$.
And yet, to know the spin connection that builds the new covariant derivative, the torsionless Maurer-Cartan equation ($d\theta^{A}+{{\omega}^{A}}_{B}\wedge\theta^{B}=0$) should be solved. The non-vanishing components become:
\begin{equation}
{{\omega_{\theta}}^{1}}_2=-{{\omega_{\theta}}^{2}}_1=-1, \quad \textrm{for} \quad r \ge R; \quad \textrm{and} \quad 
{{\omega_{\theta}}^{1}}_2=-{{\omega_{\theta}}^{2}}_1=+1, \quad \textrm{for} \quad r \le R.
\label{spin connections}
\end{equation}

Remembering that all of the arrangements build distinct from the usual six-carbon edges cell in graphene sheets are called topological defects \cite{Katsnelson}. And more, each of these defects has associated a fictitious gauge flux in the continuous regime \cite{Fialkovsky Vassilevich}. So, this work intends to describe the massless fermions by a Weyl equation in the background of a gauge field ($A_{\mu}$) bring about by the presence of heptagon rings in the wormhole nanotube junctions:  
\begin{equation}
iv_f \sigma^\mu ( \nabla_\mu - i A_\mu ) \Psi =0.
\label{Weyl equation}
\end{equation}
Where $\nabla_{\mu}=\partial_{\mu}+\Gamma_{\mu}(x)$ is the covariant derivative, with spinorial connection written by $\Gamma_{\mu}=\frac{1}{8}\omega_{\mu ab}[\sigma^{a}, \sigma^{b}]$. Using the spin connections found in (\ref{spin connections}) , the components of spinorial connections become:

\begin{equation}
\Gamma_\theta=-\frac{i\sigma^3}{2}, \quad \textrm{for} \quad r \ge R; \quad \textrm{and} \quad
\Gamma_\theta=+\frac{i\sigma^3}{2}, \quad \textrm{for} \quad r \le R.
\end{equation}

A further point to consider would be that the gauge field ($A_{\mu}$) can be written by a flux ($\Phi$) from the global contribution of each heptagon ring on the wormhole junction. So, the gauge field follow as: 
\begin{equation}
A_\theta = \pm \frac{\Phi}{2\pi}.
\label{flux of global contribution}
\end{equation}

Futhermore, the flux in (\ref{flux of global contribution}) can be indexed by region of the wormhole throat. $\tilde{\Phi}$ and $\bar{\Phi}$ for the regions inside and outside, respectively. For that we must to consider ($\tilde{\Phi}=-\bar{\Phi}$) to obtain well-behaved solutions over the entire background geometry.
Another concern might be that how bring up the Fermi points and the sub-lattice index. Particularly the following ansatz has been chosen:
\begin{equation}
\Psi=e^{-i\frac{Et}{\hbar}}\left(\begin{array}{cccc}\Psi^{\nu}_{A}  \\ \Psi^{\nu}_{B} \end{array}\right), \qquad \textrm{where} \quad \nu=\pm 1.
\end{equation}
Using that, the equation (\ref{Weyl equation}) for massless fermions results as: 
\begin{equation}
\frac{\sigma^0}{v_f}\frac{\partial}{\partial t}\left(\begin{array}{cccc}\psi^\nu_A \\ \psi^\nu_B \end{array}\right)+
\sigma^{1} \bigg (\frac{\partial}{\partial r}+\frac{1}{2r} \bigg )\left(\begin{array}{cccc}\psi^\nu_A \\ \psi^\nu_B \end{array}\right)+
\frac{\sigma^2}{r}\bigg ( \frac{\partial}{\partial \theta} + \frac{i\nu\bar{\Phi}}{2\pi} \bigg )\left(\begin{array}{cccc}\psi^\nu_A \\ \psi^\nu_B \end{array}\right)=0; \qquad r \ge R,
\label{Weyl outside}
\end{equation}
and
\begin{equation}
\frac{\sigma^0}{v_f}\frac{\partial}{\partial t}\left(\begin{array}{cccc}\psi^\nu_A \\ \psi^\nu_B \end{array}\right)+
\frac{\sigma^1 r^2}{R^2} \bigg (\frac{\partial}{\partial r}-\frac{1}{2r} \bigg )\left(\begin{array}{cccc}\psi^\nu_A \\ \psi^\nu_B \end{array}\right)+
\frac{\sigma^2 r}{R^2}\bigg ( \frac{\partial}{\partial \theta} + \frac{i\nu\tilde{\Phi}}{2\pi} \bigg )\left(\begin{array}{cccc}\psi^\nu_A \\ \psi^\nu_B \end{array}\right)=0; \qquad r \le R.
\label{Weyl inside}
\end{equation}

Note that the possibility to split the effective flux ($\Phi$) comes from the previous concern in establish a large differente between radius and wormhole throat. In this way avoiding soft curvature effects as far as possible.

\section{Quantum Holonomy}\label{secIII}

Now the possibility to realize a quantum gate based by a graphene wormhole geometric description (Eq. \ref{geometric description}) will be discused. For that, a way of obtaining the geometrical phase ($\zeta(r,\bar{\Phi})$) of the system will be established using the Weyl equation outside and inside of the throat (Eq. \ref{Weyl outside} and \ref{Weyl inside}). To do this, we must use the Dirac phase factor method, where we suppose that the Dirac spinor is written as:
\begin{equation}
\Psi^{\nu}(t, r, \theta)=e^{\zeta (r,\bar{\Phi})}\Psi^{\nu}_{0}(t, r, \theta) = \exp{ \bigg ( - \int \Gamma_{\mu}(x)dx^{\mu} \bigg )}\Psi^{\nu}_{0}(t, r, \theta)
\end{equation}
which $\Gamma_{\mu}(x)=\frac{i}{4}\omega_{\mu a b}(x)\Sigma^{ab}$ being the spin connection. 
Actually, the main reason to find the spinorial connection is the partial derivative in Weyl equation is no longer allow to assure the gauge invariance in the Lorentz group. So, local Lorentz transformations must to be introduced in the fermion couplig. Thus, we can calculate the spinorial connection by defining a local reference frame at each point along the closed curse around the defect. And so on, the holonomy matrix: $U(r, \bar{\Phi})=e^{\zeta(r,\bar{\Phi})}$, stands here for a parallel transport of a spinor along a path around (inside and outside) the throat of the wormhole. The geometrical phase in both chart is given by: 
\begin{equation}
\zeta(r,\bar{\Phi})_{in}=\frac{\ln(r)}{2\pi}(\pi-\nu\bar{\Phi}\sigma^3); \qquad r \le R,
\end{equation}
and
\begin{equation}
\zeta(r,\bar{\Phi})_{out}=\frac{\ln(r^{-2})}{4\pi}(\pi-\nu\bar{\Phi}\sigma^3); \qquad r \ge R.
\end{equation}
As you can see, both written in terms of effective flux out of throat ($\bar{\Phi}$). Which will be clarified soon.
In particular, the propertie: $[\pi, \sigma^{3}]=0$, associated with Haussdorf formula:
\begin{equation}
\exp(A)\exp(B)=\exp(A+B+{\frac {1}{2}}[A,B]+{\frac {1}{12}}([A,[A,B]]+[B,[B,A]])+...),
\end{equation}
become usefull to find the holonomic matrix as:
\begin{equation}
U(r,\bar{\Phi})_{in}=\sqrt{r} e^{-\frac{\nu\bar{\Phi}}{2\pi}\ln(r)\sigma^3}, \quad \textrm{for} \quad r \le R;
\end{equation}
and
\begin{equation}
U(r,\bar{\Phi})_{out}=\frac{1}{\sqrt{r}} e^{\frac{\nu\bar{\Phi}}{2\pi}\ln(r)\sigma^3}, \quad \textrm{for} \quad r \ge R.
\end{equation}

Here it's possible highlight that matrices are unitaries ($U_{in}=(U_{out})^{-1}$). As well, expandind as $e^A \approx 1+A+\frac{A^2}{2!}+\dots$, and doing $r\to \frac{r}{r_0}$, its possible written them as: 
\begin{equation}
U(r,\bar{\Phi})_{in}\approx \sqrt{\frac{r}{r_0}}\bigg (1-\frac{\nu}{2}\ln\bigg (\frac{r}{r_0}\bigg ) \frac{\bar{\Phi}}{\pi}\sigma^3 \bigg), \quad \textrm{for} \quad r \le R; 
\end{equation}
and
\begin{equation}
U(r,\bar{\Phi})_{out}\approx \sqrt{\frac{r_0}{r}}\bigg (1+\frac{\nu}{2}\ln\bigg (\frac{r}{r_0}\bigg ) \frac{\bar{\Phi}}{\pi}\sigma^3 \bigg), \quad \textrm{for} \quad r \ge R.
\end{equation}

As it might be seen, running $r\to r_{0}$, the consequence is $U_{in}=U_{out}=1$.
Namely, is very important result because quantum gates should be built over any background as unitary transformations. This allow the Hamiltonian in the fermion couplig gives as look forward by an unitary time evolution of the system. And, as would be expected, the function $U(r,\bar{\Phi})$ still regular everywhere. Another consideration would be that the system have $\bar{\Phi}$ as the main control parameter. What on one hand clarify all the enginnering behind of system over just the electronic density of states around the wormhole, on other slight the range of possibilities to work on over that. 

\section{Conclusion}\label{secIV}

In conclusion, a discussion over a possible way to build quantum holonomy over graphene wormholes has been brought up. The quantum fermion coupling by an effective field theory has been investigated inside a geometrical model for graphene wormhole proposed in \cite{Gonzalez Herrero}. Notably, the topological defect's contribution over the holonomic matrix could sum up to an effective flux out of the throat. Considering this region is the only one with the practical ground for that. In addition, the radius of the throat was considered much larger than its length to avoid curvature effects. All this made it possible to infer a holonomic matrix that supports futures works over quantum gates into wormholes backgrounds in bidimensional materials.
Another point to consider might be the possible existence of graphene wormholes with different distributions of defects \cite{Gonzalez Guinea Herrero, Kleinert Gauge Fields in Condensed Matter}. Changing the localized states and the effective flux ($A_{\mu}$) in the fermion coupling. In that situation, we wish to be possible still work with just one gauge field. Or, at least, a couple of that going through the two-dimensional graphene lattice. 

Furthermore, the change of wave function of the electron troughtout the throat could be seen as a phase obtained by a Cooper pair, in a kind of Josephson-junction. The electron crossing between the graphene sheets would be exatcly a Josephson current \cite{Superconductivity in carbon nanotubes}. Some works already pointed up high temperatures ($T \approx 12 K$) superconductivities behavior in specific cases \cite{Joseph junctions in graphene, Superconductivity in carbon nanotubes at 12K, Superconductivity em duas monocamadas}.
Moreover, that tunel efect might be seen as a flip of the qubit 0 to 1 in logical base preconditionated. Enlarging the range of possibilities to work quantum computation over these systems.

\section{ACKNOWLEDGMENTS}

$\quad$I thank CAPES and CNPQ for financial support.

\section{Data Availability Statement}


Data sharing not applicable to this article as no datasets were generated or analysed during the current study.

\end{document}